\documentclass[11pt,twoside]{article}
\usepackage{asp2004}
\usepackage{psfig}
\usepackage{epsf}
\usepackage{graphics}
\usepackage{lscape}
\def\edcomment#1{\iffalse\marginpar{\raggedright\sl#1\/}\else\relax\fi}
\reversemarginpar

\begin{document}
\title{{\it XMM-Newton} detects the beginning of the X-ray decline of SN 1995N}
\author{P. Mucciarelli}
\affil{INAF-Astronomical Observatory of Padova \\ 
Department of Astronomy, University of Padova}
\author{L. Zampieri}
\affil{INAF-Astronomical Observatory of Padova}
\author{A. Pastorello}
\affil{INAF-Astronomical Observatory of Arcetri\\
Max Planck Institut f\"ur Astrophysik, Garching}

\begin{abstract}
We present the results of a new {\it XMM-Newton} observation of the 
interacting supernova 1995N, performed on July 27, 2003. We find that 
the 0.2-10.0 keV flux has dropt at a level of $1.44 \times 10^{-13}$ 
erg cm$^{-2}$ s$^{-1}$, about one order of magnitude lower than that 
of a previous {\it ASCA} observation performed on January 1998. The X-ray spectral 
analysis shows statistically significant evidence for the presence 
of two distinct components, that can be modeled with emission from 
optically thin, thermal plasmas at different temperatures. From these 
temperatures we derive that the exponent of the ejecta density 
distribution is $n \sim 6.5$.
\end{abstract}

\thispagestyle{plain}

\noindent{SN 1995N, discovered on May 1995 several months after the 
explosion \citep{benetti95}, is of special interest in the context 
of circumstellar medium (CSM) interacting supernovae. It is hosted 
in the (IB(s)m pec) galaxy MCG-02-38-017, at a distance of $\sim$ 24 Mpc. 
The epoch of explosion is not known, but was estimated to
be about 10 months before its optical discovery \citep{benetti95}.} 

\section{{\it XMM-Newton} observation of SN 1995N}
\begin{figure}[ht!]
\begin{center}
\scalebox{0.25}{\rotatebox{-90}{\includegraphics{mucciarelli_fig1.eps}}}
\scalebox{0.26}{\rotatebox{-90}{\includegraphics{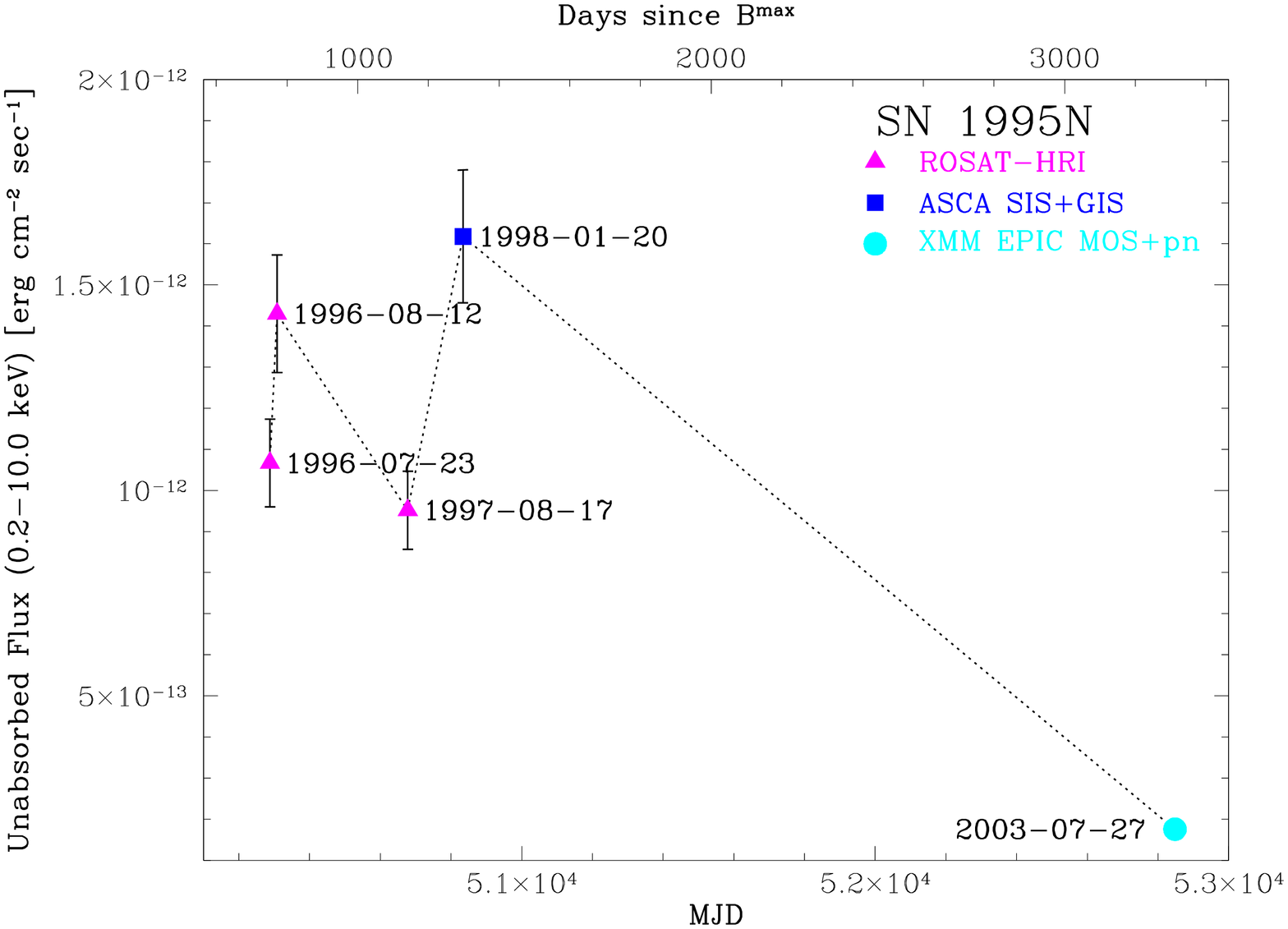}}}
\caption{ {\it Left panel}: 
{\it XMM} EPIC MOS and pn spectra of SN 1995N 
in the 0.2-10.0 keV interval, along with the best fitting continuum model
(solid line) and the two {\sc mekal} components (dotted lines). 
{\it Right panel}: 
0.2-10.0 keV unabsorbed lightcurve of SN 1995N. Fluxes were de-absorbed and, 
when necessary, extrapolated in the 0.2-10.0 keV interval. An estimate of 
the errors, based on counting statistics, is $\sim$ 10\%.} 
\label{fig1}
\end{center}
\end{figure}

We observed SN 1995N with {\it XMM-Newton} for 72 ks on July 27, 2003,
nine years after the explosion. Here we present a preliminary analysis 
of this observation \citep[for details see][hereafter Z04]{zamp04}. The 
X-ray pointing was coordinated with near-IR and optical observations 
\citep[see][]{pasto04}. The X-ray data set was heavily affected by solar 
flares. We analysed the longer {\it XMM} EPIC exposures of both the MOS 
and pn detectors (56 and 64 ks respectively). Data were filtered using 
the count rate criterion $<$ 0.5 counts s$^{-1}$ for the EPIC MOS and 
$<$ 1.0 counts s$^{-1}$ for the EPIC pn, leaving 22 and 14 ks of useful 
data, respectively. Source counts were estracted from a circular region 
of 20$\arcsec$ centered on the radio position of the supernova \citep{vd96}. 
Background counts were estracted from a circular region of 40$\arcsec$, 
on the same CCD. The source count rate was 9.7$\times 10^{-3}$ counts 
s$^{-1}$ for the EPIC MOS ($\sim$450 counts) and 3.3 $\times 10^{-2}$ 
counts s$^{-1}$ for the EPIC pn ($\sim$500 counts). 
EPIC MOS and pn spectra were binned requiring at least 15 counts per bin. 
Joint MOS and pn spectral fits were performed in the 0.2-10.0 keV interval 
with an overall normalization constant. Despite the low statistics, the fit 
with single component models was not satisfactory (see Z04). 
The improvement obtained adding a {\sc mekal}\footnote{The {\sc mekal} model 
is the spectrum emitted by an optically thin, thermal plasma.} component 
to a single component spectral models was significant at the $\sim$4$\sigma$ 
level. The best fit was obtained with an absorbed double {\sc mekal} model 
with column density of the interstellar medium $N_H = 1.3 \times 10^{21}$ 
cm$^{-2}$ and temperatures of the two thermal components $kT\simeq$ 0.8 
and 9.5 keV ($\chi^2_{red} \simeq 0.76$; see Figure \ref{fig1}, 
{\it left panel}). 

We also report the detection of a faint object inside the X-ray error box 
of SN 1995N in the summed {\it XMM} OM $UVW1$ image. The object was at the 
limit of detectability in each single frame. Its position in the summed 
image is within 0.4$''$ from the radio position of SN 1995N. Within the 
errors of the astrometric calibration, we then identify this object with 
SN 1995N. The lack of reference stars in the $UVW1$ band prevented us from 
performing a photometric calibration of the image. Taking the detection limit 
count rate of a single image as a lower bound for the UVW1 flux of SN 1995N, 
we estimate $F_{UVW1}\ga 6.3 \times 10^{-15}$ erg cm$^{-2}$ s$^{-1}$.

\section{X-ray variability}\label{var}

Figure \ref{fig1} ({\it right panel}) shows the fluxes derived  from all 
the available X-ray observations of SN 1995N. The first X-ray observation 
was performed with {\it ROSAT} HRI on July 23, 1996 \citep*[1.3 ks; ][]{lewin96}, 
followed by other two exposures taken on August 12, 1996 (17 ks) and on 
August 17, 1997 (19 ks). The {\it ROSAT} fluxes were derived from the count 
rates reported by \citet{fox00}, assuming a {\sc power-law} spectrum with 
photon index and column density from their fit of the {\it ASCA} spectra. 
The {\it ASCA} observation was performed on January 19, 1998 
\citep[83 and 96 ks, respectively for the SIS and GIS instruments;][]{fox00}. 
For {\it ASCA} and {\it XMM}, the fluxes reported in Figure \ref{fig1} are 
the average between the different instruments. It is worth noting the large 
decrease in the X-ray flux between the {\it ASCA} and {\it XMM} observations 
(unabsorbed [0.2-10.0 keV] fluxes of 1.65 $\times 10^{-12}$ and 
1.75 $\times 10^{-13}$ erg cm$^{-2}$ s$^{-1}$, respectively).

\section{Discussion}
SN 1995N is one of the few supernovae detected in X-rays at an age of
$\sim$9 years. Our {\it XMM} observation shows that the unabsorbed X-ray 
flux of SN 1995N dropped at a value of $1.75 \times 10^{-13}$ erg 
cm$^{-2}$ s$^{-1}$, about an order of magnitude lower than that of the 
previous {\it ASCA} observation performed $\sim$6 years before. The 
decline of the X-ray flux is signalling that SN 1995N probably started to evolve 
towards the remnant stage. Interpreting the evolution of the X-ray light 
curve is not straigthforward. A complex scenario in which a two-phase 
(clumpy and smooth) CSM contributes to the observed X-ray emission is 
consistent with the available data (see Z04).

The EPIC spectrum of SN 1995N shows statistically significant evidence
for the presence of two distinct thermal (MEKAL) components at temperatures 
of 0.8 and 9.5 keV, respectively. In the standard model of ejecta/wind 
interaction \citep{cf94}, these represent the temperatures of the gas 
between the contact discontinuity and the reverse/forward shock. The 
temperature of the hotter phase is similar to the temperature of the 
single-component spectral fit of the {\it ASCA} data performed by \cite{fox00}. 
Within the assumptions of the standard  model, we can derive the exponent 
of the ejecta density distribution $n$ from the expression 
$T_1/T_2 = (3-s)^2/(n-3)^2$, where $s$ is the exponent of the CMS density 
distribution \citep*{fran96}. Assuming a constant and homogeneous stellar 
wind ($s=2$), from the values of the temperatures $T_1$ and $T_2$ inferred 
from the X-ray spectrum, we obtain $n\sim6.5$.

\end{document}